\documentclass[10pt,conference,a4paper,twoside,twocolumn]{IEEEtran}
\usepackage{amsthm}
\usepackage{amsmath,amssymb}
\usepackage{psfrag}
\usepackage{pifont}
\usepackage{graphicx}
\usepackage[nocompress]{cite}
\graphicspath{{drawings/}{pics/}}
\usepackage{color}
\usepackage{array}
\usepackage[normalem]{ulem}
\usepackage{multirow}

\newcommand{\const}{\mathcal{X}}
\newcommand{\map}{\mathcal{M}}
\newcommand{\CM}{^{\mathrm{CM}}}
\newcommand{\BICM}{^{\mathrm{BICM}}}
\newcommand{\Cconst}{{C\CM_{\hspace*{-.4mm}\const}}}
\newcommand{\Cbicm}{C\BICM_{\hspace*{-.4mm}\const,\map}}
\newcommand{\slopelin}{s_0}
\newcommand{\slopelog}{\mathcal{S}_0}
\newcommand{\cone}{{c_1^{}}}
\newcommand{\ctwo}{{c_2^{}}}
\newcommand{\coneCM}{{c_1\CM}}
\newcommand{\ctwoCM}{{c_2\CM}}
\newcommand{\coneBICM}{{c_1\BICM}}
\newcommand{\ctwoBICM}{{c_2\BICM}}
\newcommand{\mean}[1]{\ve{m}_{\hspace*{-.4mm}#1}}
\newcommand{\cov}[1]{\ve{M}_{\hspace*{-.7mm}#1}}
\newcommand{\EbNolim}{(\Eb/\No)_{\mathrm{lim}}}

\usepackage[active]{srcltx}

\def\ve#1{{\mathchoice{\mbox{\boldmath$\displaystyle #1$}}%
              {\mbox{\boldmath$\textstyle #1$}}%
              {\mbox{\boldmath$\scriptstyle #1$}}%
              {\mbox{\boldmath$\scriptscriptstyle #1$}}}}
\DeclareSymbolFont{AMSb}{U}{msb}{m}{n}
\DeclareSymbolFontAlphabet{\mathbb}{AMSb}

\def\R{\mathbb{R}}

\def\trace{\mathrm{trace}}
\def\e{\mathrm{ e}}

\def\trans{^\mathsf{T}}

\def\E{\mathrm{E}}

\newcommand {\deq} {\stackrel{\mathrm{def}}{=}}     
\newcommand {\dx} {~\mathrm{d}}

\newcommand {\Es} {E_{\mathrm{s}}}            

\def\dB{\mathrm{dB}}

\newcommand {\Eb} {E_{\mathrm{b}}}     
\def\No{N_0}

\DeclareFontFamily{OT1}{phv}{}
\DeclareFontShape{OT1}{phv}{m}{n}{ <-> [0.95] phvr8t }{}

\title{\huge Capacity of BICM Using (Bi-)Orthogonal Signal Constellations in Impulse-Radio Ultra-Wideband Systems}

\IEEEoverridecommandlockouts
\author{
\IEEEauthorblockN{Andreas~Schenk and Robert~F.H.~Fischer}
\IEEEauthorblockA{Lehrstuhl f{\"u}r Informations{\"u}bertragung, Friedrich-Alexander-Universit{\"a}t Erlangen--N{\"u}rnberg, Erlangen, Germany}
Email: \texttt{\{schenk,fischer\}@lnt.de}
\thanks{This work was supported by the Deutsche Forschungsgemeinschaft (DFG) within the framework UKoLoS under grant FI 982/3-1/2.}
}
\begin{document}

\setlength{\textheight}{1.022\textheight}
\maketitle
\begin{abstract}
Bit-interleaved coded modulation (BICM) using \mbox{(bi-)}orthogonal signals is especially well suited for the application in impulse-radio ultra-wideband transmission systems, which typically operate in the power-limited regime and require a very low-complexity transmitter and receiver design.
In this paper, we analyze the capacity of BICM using (bi-)orthogonal signals with coherent and noncoherent detection and put particular focus on the power-limited or so-called wideband regime.
We give analytical expressions for the ratio energy per bit vs.\ noise power spectral density in the limit of infinite bandwidth and the respective  wideband slope, and thus, are able to quantify the loss incurred by the restriction to BICM in contrast to coded modulation.
The gained theoretical insights allow to derive design rules for impulse-radio ultra-wideband transmission systems.

\end{abstract}

%
\section{Introduction}
\label{sec:intro}

Very low-complexity signaling schemes, such as pulse-position modulation,
have seen a revival, in particular in the emerging field of impulse-radio ultra-wideband (IR-UWB)
transmission systems \cite{UWB:Aiello:UWB,UWB:Giannakis:UWB}, which are widely regarded as a promising technique for short-range, low-data rate
applications like, e.g., wireless sensor networks \cite{UWB:Zhangetal:UWBWirelessSensorNetworks}.
There are two main reasons why (bi-)orthogonal signaling schemes, such as
pulse-position modulation (PPM) or its biorthogonal extension (biPPM), are employed in IR-UWB systems \cite{UWB:ZhangGulliver:BiorthogonalPPMTHUWBIR}.
First, hardware restrictions on transmitter and receiver complexity as well as direct transmission of
the IR-UWB transmit signal without upconversion to a carrier frequency prohibit the use of higher-order
amplitude or even quadrature modulation \cite{UWB:Aiello:UWB,UWB:Win:IR}.
Second, due to the large signal bandwidth, IR-UWB systems typically operate in the power-limited or so-called wideband regime \cite{IT:Verdu:SpectralEfficiency}.

IR-UWB, however, is only one---but maybe the most prominent---application example.
Orthogonal schemes are, e.g., applied in free space optical communications \cite{BICM:NguyenLampe:MPPMFreespaceOptics} and as on/off frequency-shift keying in wideband fading channels \cite{IT:GursoyPoorVerdu:OnOffFSKforWidebandFading}.
Apart from this, orthogonal schemes are of particular theoretical relevance, cf., e.g., \cite{IT:Gursoy:EnergyEfficiencyofOrthogonalSignaling,IT:Verdu:SpectralEfficiency}.

To keep transmitter and receiver design simple, in contrast to coded modulation (CM) employing, e.g., multi-level codes \cite{IT:WachsmannFischerHuber:Multilevelcodes},
IR-UWB systems often restrict to the conventional serial concatenation of coding and modulation at transmitter, and
detection and decoding at receiver side, i.e., restrain to the bit-interleaved coded modulation (BICM) philosophy \cite{BICM:Caire:BICM}.
BICM has most commonly been studied for carrier modulated digital quadrature
amplitude modulation, cf., e.g., \cite{BICM:Caire:BICM,BICM:Alvarado:Diss}, and in particular \cite{BICM:MartinezFabregasCaireWillems:BICMWideband} for an analysis of BICM in the wideband regime.
From this extensive analysis it is well known that the employed binary labeling of the signal elements
has significant influence on the BICM capacity especially in the wideband regime, cf., e.g., \cite{BICM:ClemensRobert:MappingsBICMUWB,BICM:ClemensRobert:AsymptoticallyOptimalMappings}.
BICM of orthogonal signal schemes has been assessed briefly in \cite{BICM:Caire:BICM} and more detailed in \cite{BICM:ChengValenti:TurboNoncoherentOrthogonal} for iterative decoding.

Motivated by IR-UWB as a field of application, in this paper we analyze BICM using (bi-)orthogonal signaling in the wideband regime \cite{IT:Verdu:SpectralEfficiency,BICM:MartinezFabregasCaireWillems:BICMWideband}.
We give a detailed analysis of the BICM capacity (including the wideband slope and the asymptotic energy per bit vs.\ noise spectral density in the limit of infinite bandwidth) using \mbox{(bi-)}orthogonal signaling for the case of coherent detection, present a (near-)Gray binary labeling rule for biPPM, and compare the results with the case of noncoherent detection, which is of particular interest for IR-UWB communications \cite{UWB:Witrisaletal:NoncoherentUWBSystems}.
The gained insights in BICM of (bi-)orthogonal signaling allow to derive design
rules for IR-UWB systems.

The paper\footnote{
Detailed derivations can be found in the appendix of an extended version of the paper on arXiv (arXiv:1102.2761).
} is outlined as follows:
After introducing the system model and its relation to IR-UWB in Sec.\,\ref{sec:constellation}
and a brief review of BICM and the analysis in the wideband regime in Sec.\,\ref{sec:BICM},
the main results on BICM using \mbox{(bi-)}orthogonal signal constellations are presented in Sec.\,\ref{sec:BICMbiPPM} and visualized in Sec.\,\ref{sec:results}.
The paper concludes with a summary of the results and their consequences on the design of IR-UWB systems in Sec.\,\ref{sec:conclusions}.

%
\section{IR-UWB and BICM}
\label{sec:constellation}
In IR-UWB systems, impulses of very short duration in the order of nanoseconds
are used to generate the transmit signal directly in the baseband.
This very low-complexity transmitter design in combination with baseband transmission
enables to only encode information bits in the pulse position and/or by inverting the pulse amplitude.
To avoid inter-symbol interference even in dense multipath propagation scenarios
the spacing of the pulses in time domain
is chosen large enough to ensure each received pulse has decayed before the next
pulse is received.
Denoting the pulse spacing by $\Delta$, each symbol occupies a time slot of $T =D \Delta$, where $D$ denotes the number of possible pulse positions per symbol (dimension).
Note that in general each symbol may be represented by more than one pulse, to enable time-hopping and/or code division multiple-access \cite{UWB:Giannakis:UWB}.

Employing, e.g., a RAKE-receiver, in the case of coherent detection of IR-UWB, this motivates to use the following $D$-dimensional real-valued discrete-time equivalent signal model, where the receive vector is given as%
\footnote{Boldface letters denote (row-)vectors, upper-case letters denote random variables, and lower-case letters the particular realization (only exception: cov.\ matrix $\cov{\hspace*{.3mm}\cdot}$ and identity matrix $\ve{I}$). $I(\cdot;\cdot)$: mutual information, $\E\{\cdot\}$: expectation operator, 
$[\cdot]_{r,c}$: element in row $r$ and column $c$ of a matrix.}
\begin{align}
  \ve{y} = \ve{x} + \ve{n} \label{eq:model}
\end{align}
where $\ve{n}$ is the additive noise vector comprised of uncorrelated Gaussian distributed entries,
each with variance $\sigma_n^2 = \No/2$, and $\ve{x}$ is the transmitted signal element drawn  from the real-valued $D$-dimensional signal set $\const\subseteq\R^D$.
After suitable normalization, the energy per symbol is given as $\Es=1$.
We denote the cardinality of the signal set by $|\const| = M$ and by $m = \log_2(M)$ the number of bits required to address a signal element.
The mapping from binary $m$-tuples $\ve{b} = [b_1,\ldots ,b_{m}] \in\{0,\,1\}^m$ to the signal elements $\ve{x}\in\const$ is specified by a bijective binary labeling rule $\map: \ve{b}\mapsto\ve{x}$.
We assume the binary labels to be chosen equiprobably.
For future use we define 
the mean $\mean{\const}$ and the covariance matrix $\cov{\const}$ of the signal set, respectively as 
\begin{align}
   \mean{\const} &= \E\left\{ \ve{X} \right\}\label{eq:mean}\\
%
   \cov{\const} &=  \E\left\{  (\ve{X}-\mean{\const})\trans(\ve{X}-\mean{\const}) \right\}\\
		&= \E\{\ve{X}\trans\ve{X}\} - \mean{\const}\trans\mean{\const}\;.\label{eq:var}
\end{align}
In Sec.\,\ref{sec:BICMbiPPM} we focus on (bi-)orthogonal signal constellations.

%
\section{Review of Bit-Interleaved Coded Modulation}
\label{sec:BICM}
\subsection{Capacity of BICM}
According to \cite{IT:WachsmannFischerHuber:Multilevelcodes}, the coded modulation (CM), or constellation constrained, capacity $\Cconst$
is given as the mutual information between the channel input and channel output
\begin{align}
   \Cconst  = I(\ve{X};\ve{Y}) &= I(B_1B_2\ldots B_m;\ve{Y}) \label{eq:cm_cap}\\
            &= I(B_1;\ve{Y}) + I(B_2;\ve{Y}|B_1) + \ldots \nonumber\\
            &\qquad + I(B_m;\ve{Y}|B_1B_2\ldots B_{m-1}) \;.\nonumber
\end{align}
For brevity the dependence on the ratio $\Es/\No$ is not indicated explicitly.
The second line follows from the chain rule of information theory and can be interpreted as the parallel transmission of the binary label entries $b_i$, $i=1,\ldots ,m$, over $m$ memoryless binary input channels, followed by successive decoding \cite{IT:WachsmannFischerHuber:Multilevelcodes}, i.e., for decoding of a particular bit level the knowledge of the lower bit levels is taken into account.

Neglecting the contribution of the lower bit positions in the decoding process, i.e., if parallel decoding is performed, gives the BICM capacity (with parallel decoding) as the sum of the bit level capacities $I(B_i;\ve{Y})$ \cite{BICM:Caire:BICM}.
Equivalently, applying the chain rule, we have (cf.\ \cite{BICM:MartinezFabregasCaireWillems:BICMWideband})
{\allowdisplaybreaks
\begin{align}
   \Cbicm &= \sum_{\mu=1}^m I(B_i;\ve{Y}) \label{eq:bicm_cap}\\
   &= \sum_{\mu=1}^m \left(I(B_0\ldots B_{i-1} B_i B_{i+1}\ldots B_m;\ve{Y})\right.\nonumber\\[-3mm]
                      &{}\quad\qquad\left.- I(B_0\ldots B_{i-1} B_{i+1}\ldots B_m;\ve{Y}|B_i)\right)\nonumber\\
                    &= \sum_{\mu=1}^m \left( \Cconst - \frac{1}{2}\sum_{b\in\{0,1\}} C^{\mathrm{CM}}_{\hspace*{-.4mm}\const_b^{\mu}} \right) \label{eq:bicm2_cap}
\end{align}}
where
\begin{align}
C^{\mathrm{CM}}_{\hspace*{-.4mm}\const_b^{\mu}} &= I(B_0\ldots B_{i-1} B_{i+1}\ldots B_m;\ve{Y}|B_i=b)\\
               &= I(\ve{X};\ve{Y}|\ve{X}\in\const_b^{\mu}) \label{eq:const_cap}
\end{align}
is the CM capacity of the constrained signal set
\begin{align}
   \const_b^{\mu} = \{\ve{x}\, |\,& \ve{x}=\map([b_1\ldots b_m]), b_\mu = b, \label{eq:const_const}\\
              & b_{\nu} \in \{0,1\},\,\forall \nu=1,\ldots,m, \nu\neq \mu\} \nonumber
\end{align}
composed of all signal elements corresponding to the bit label $b$ at position $\mu$.
Eq. (\ref{eq:bicm2_cap}) gives a direct relation between the CM capacity and the BICM capacity.
As indicated, in contrast to $\Cconst$, $\Cbicm$ depends on the binary labeling rule $\map$ \cite{BICM:Caire:BICM}.

\subsection{Analysis in the Wideband Regime}
The analysis in the wideband regime, i.e., $C\rightarrow0$, as defined in \cite{IT:Verdu:SpectralEfficiency}, is equivalent to the analysis of the capacity in the limit of vanishing $\Es/\No \rightarrow 0$.
It is then of particular interest to derive the corresponding ratio $\EbNolim$ of the capacity curve as a function of $\Eb/\No$, where $\Eb$ denotes the energy per information bit, i.e., $\Eb/\No = \Es/\No / C$,  and the slope at this point \cite{IT:Verdu:SpectralEfficiency}.

To this end, applying a second-order Taylor-series expansion at $\Es/\No =0$ yields (briefly sketched following \cite{BICM:ClemensRobert:MappingsBICMUWB})
\begin{align}
   C(\Es/\No) = (\cone(\Es/\No)  + \ctwo(\Es/\No)^2
)/\log(2) \;.
\end{align}
The coefficients $\cone$ and $\ctwo$ depend on the chosen signal set and the receiver design (in particular, CM or BICM).
Rewriting the capacity as a function of $\Eb/\No$, leads to
\begin{align}
   C(\Eb/\No) = \slopelin \left(\Eb/\No - \EbNolim\right)
\end{align}
which gives a linear expansion of the capacity in bits as a function of $\Eb/\No$ at the point $\EbNolim$, where
\begin{align}
   \EbNolim &= \log(2) / \cone \label{eq:ebnolim} \;.
\end{align}
The corresponding slope $\slopelin$, i.e., the so-called wideband slope (in linear scale), is given by
\begin{align}
 \slopelin &= -\frac{{\cone}^3}{\ctwo\log^2(2)} \label{eq:slopelin}\;.
\end{align}
It often provides more insight, to analyze the capacity normalized to the dimension of the signal constellation and directly as a function of $10\log_{10}(\Eb/\No)$, i.e., in decibels \cite{IT:Verdu:SpectralEfficiency};
the wideband slope (in bit/dimension per $3\,\dB$) is then given as (note the factor $D$ compared to \cite{IT:Verdu:SpectralEfficiency} due to the signaling in $D$ real dimensions)
\begin{align}
 \slopelog &= -\frac{\cone^2}{D\ctwo} \label{eq:slopelog}\;.
\end{align}

From Theorem\,5 in \cite{IT:PrelovVerdu:SecondOrderAsymptotics} the coefficients of the CM capacity $\Cconst$ around $\Es/\No=0$ for coded modulation schemes and multidimensionsal real-valued signal sets are given as
\begin{align}
   \coneCM &= \trace \left(\cov{\const}\right),\quad
   \ctwoCM = - \trace \left(\cov{\const}^2\right) \label{eq:c2_cm_trace}\;.
\end{align}
Noting that $\const$ is normalized to unit energy, with (\ref{eq:var}) we have
\begin{align}
   \coneCM &= 1-\|\mean{\const}\|^2 \label{eq:c1_cm}\\
   \ctwoCM &= - \trace \left( \right( \E\{\ve{X}\trans\ve{X}\} - \mean{\const}\trans\mean{\const}  \left)^2\right) \label{eq:c2_cm}\;.
\end{align}

With (\ref{eq:ebnolim}), (\ref{eq:slopelog}), signaling with $\const$ is only wideband-optimal \cite{IT:Verdu:SpectralEfficiency} (i.e., the fundamental limit $\EbNolim = \log(2)$ is approached at a slope of $1$ bit/dimension per $3\,\dB$), if $\cone = 1$ and $\ctwo = -1/D$.
This translates to i) $\const$ has zero-mean ($\|\mean{\const}\|=0$) and ii) $\const$ is a proper constellation, i.e., $\E\{\ve{X}\trans\ve{X}\} = \ve{I}/D$, meaning that the dimensions are uncorrelated and the energy is equally spread among the dimensions \cite{BICM:MartinezFabregasCaireWillems:BICMWideband}.
The latter directly follows as the maximum of the trace of $(\E\{\ve{X}\trans\ve{X}\})^2$ is then attained (subject to $\E\{\|\ve{X}\|^2\} = 1$).

Using a slight generalization of Theorem\,2 of \cite{BICM:MartinezFabregasCaireWillems:BICMWideband}, (combining (\ref{eq:bicm2_cap}), (\ref{eq:c2_cm})), the coefficients of the BICM capacity read
\begin{align}
   \coneBICM&= \frac{1}{2}\sum_{\mu=1}^m \sum_{b=\{0,1\}} \left(\trace \left(\cov{\const}\right) - \trace \left(\cov{\const_b^{\mu}}\right) \right) \raisetag{8mm}\label{eq:c1_bicm}\\
   \ctwoBICM &= \frac{1}{2}\sum_{\mu=1}^m \sum_{b=\{0,1\}} \left(-\trace \left(\cov{\const}^2\right) + \trace \left(\cov{\const_b^{\mu}}^2\right)\right) \raisetag{3mm}\label{eq:c2_bicm}
\end{align}\\[-3mm]
where $\cov{\const_b^{\mu}}$ is the covariance matrix of the constrained constellation $\const_b^{\mu}$ (cf.\ (\ref{eq:var})).

\section{BICM using (Bi-)Orthogonal Signals}
\label{sec:BICMbiPPM}

The most simple case---both from hardware implementation complexity, as well as from theoretical perspective---is orthogonal $M$-ary pulse-position modulation ($M$-PPM).
The $(D=M)$-dimensional signal elements are given by $\const = \{ \ve{e}_i| i=1,\ldots,M\}$,
where $\ve{e}_i$ is the $i$-th unit vector.

The natural expansion of this case is biorthogonal pulse-position modulation ($M$-biPPM), i.e., the negatives of the orthogonal PPM signals are included in the signal set \cite{book:Proakis:DigitalCommunications}, yielding in total $M = 2D$ signal elements.
Of special interest is the well-known one-dimensional case of $2$-biPPM, i.e., binary phase-shift keying (BPSK).

\subsection{Orthogonal Signals ($M$-PPM)}
As the signal elements are orthogonal, $(M=D)$-ary PPM clearly froms a proper constellation, i.e., $\E\{\ve{X}\trans\ve{X}\} = \ve{I}/M$,
but it has non-zero mean $\mean{\const} = \ve{1}/M$ (with all-ones vector $\ve{1}$).
Using (\ref{eq:c1_cm}), (\ref{eq:c2_cm}), the coefficients of the CM capacity read
\begin{align}
   \coneCM = 1-1/M, \quad \ctwoCM = - \left(1-1/M\right)/M \label{eq:coeffs_cmppm}
\end{align}
and the asymptotic value and the wideband slope are given as
\begin{align}
   \EbNolim\CM = \frac{M}{M-1}\log(2),\quad
   \slopelog\CM = \frac{M-1}{M} \label{eq:vals_cmppm}\;.
\end{align}
Hence, only for $M\rightarrow \infty$, the fundamental limit of $10\log_{10}(\Eb/\No) = -1.59\,\dB$ can be achieved ($\cone \rightarrow 1$), a well-known fact for long years \cite{book:Proakis:DigitalCommunications,IT:Golay:NoteTheoreticalEfficiencyPPM}.

Due to the orthogonality of the signal elements, the Euclidean distance between any of the signal elements is equal ($\|\ve{x}_i - \ve{x}_k\| = \sqrt{2}$, $i\neq k$) \cite{book:Proakis:DigitalCommunications}.
Consequently, the mapping does not influence the BICM capacity \cite{BICM:NguyenLampe:MPPMFreespaceOptics}.
Hence, natural labeling, i.e., enumerating the signal elements according to their position index in binary representation, can be applied.
since each constrained constellation $\const_b^{\mu}$ corresponds to a $(M/2)$-(sub)PPM, independent of the particular labeling, it is easy to show
that $\trace(\cov{\const_b^{\mu}}) = 1 - 2/M$ and 
$\trace(\cov{\const_b^{\mu}}^2) = (1 - 2/M)\cdot2/M$ holds for $\mu=1,\ldots,m$, $b\in\{0,\,1\}$, thus
\begin{align}
   \coneBICM = \frac{\log_2(M)}{M}, \quad \ctwoBICM = \frac{\log_2(M)}{M} \left(1 - 3/M\right) \label{eq:coeffs_bicmppm}\;.
\end{align}
In contrast to the CM capacity, with increasing $M$ also the ratio $\EbNolim\BICM = \frac{M}{\log_2(M)}\log(2)$ increases, and, as $\ctwo>0$ for all $M\geq4$, a negative wideband slope results.

\subsection{Biorthogonal Signals ($M$-biPPM)}
Including the negatives of the signal elements of $D$-dimensional PPM, one arrives at $(M=2D)$-biPPM.
Clearly the signal set is proper, and, as opposed to $M$-PPM, has zero mean.
Thus, CM of biorthogonal signaling is wideband-optimal for all $M$, as the $\EbNolim\CM = \log(2)$ is approached at a slope of $\slopelog\CM = 1$, since $\coneCM = 1$, $\ctwoCM = -1/D$

As already mentioned, the biPPM BICM capacity, however, depends on the applied binary labeling rule.
After introducing two different labeling rules for biPPM, we derive the respective coefficients required for the wideband analysis.

\subsubsection{Binary Labeling for biPPM}
Beginning with the trivial case of $2$-biPPM ($M=2D=2$), which is equivalent to BPSK, the only two possible labeling strategies obviously lead to the same BICM capacity, which equals the CM capacity.

For higher-dimensional biPPM, a straight-forward labeling rule is given by using $m-1$ bits to specify the position index in binary representation, and an additional bit, e.g., the most-significant bit, for the sign information.
Due to the similarity to natural labeling for $M$-PPM, we denote this strategy as natural labeling for biPPM.

However, in the case of two dimensional biPPM, i.e., $4$-biPPM ($M=2D=4$), there are only two possible pulse positions, each with sign $\pm1$. Thus, the equivalence to quadrature phase-shift keying, i.e., $4$-PSK with $\pm1$ either in the real- or imaginary part, is evident.
Hence, similar to $4$-PSK, Gray labeling can be applied to $4$-biPPM, and the BICM capacity equals the CM capacity.

For higher-dimensional biPPM ($M=2D\geq8$) Gray labeling is not possible anymore.
To see this, note that each signal element has $M-2$ neighbors at distance $\|\ve{x}_i - \ve{x}_k\|=\sqrt{2}$ and only a single one, namely its counterpart, at distance $\|\ve{x}_i - (-\ve{x}_i)\|=2$.
Using the inverse labeling for the negative signal element, the remaining $m-1 = \log_2(M/2)$ bits are not sufficient to force the binary labels of the $M-2$ neighbors to be different at only a single bit position.
A straight-forward relaxation to the Gray labeling principle is to allow more than one different bit positions, i.e., applying a near-Gray labeling.
This can easily be constructed by specifying the position index in binary representation, and bitwise applying the XOR operation with an additional bit, which specifies the position.

Exemplarily, both labeling rules, near-Gray and natural labeling, are given for $8$-biPPM in Table \ref{mapping:8biPPM}.

\subsubsection{Wideband Analysis of BICM biPPM}
Computing the coefficients $\coneBICM$ and $\ctwoBICM$ to obtain $\EbNolim\BICM$ and the wideband slope in the case of BICM of biPPM, we are interested in the trace of the (squared) covariance matrices $\cov{\const_b^{\mu}}$ of the constrained constellations $\const_b^{\mu}$.

In the case of (near-)Gray labeling, each constrained constellation again corresponds to $(D=M/2)$-PPM with possibly alternating pulse amplitude.
E.g., for $8$-biPPM, fixing the second bit position to $b_2=0$ gives $\mean{\const_0^2} = 2/M\cdot\left[1,\, 1,\, -1,\, -1\right]$ (cf.\ Table \ref{mapping:8biPPM}).
Thus, one obtains (for all $i=1,\ldots,D$, $ \mu=1,\ldots,m$, $b\in\{0,\,1\}$)
\begin{align}
  \trace(\cov{\const_b^{\mu}}) &= 1-\frac{1}{D}\,,\quad
  \trace(\cov{\const_b^{\mu}}^2) = ({1-\frac{1}{D}})/{D}\nonumber
\end{align}
which gives
\begin{align}
   \coneBICM =& \frac{\log_2(2D)}{D},\quad
   \ctwoBICM = \frac{\log_2(2D)}{D} \cdot \frac{-1}{D} \label{eq:coeffs_graybippm} \\
   \EbNolim\BICM &= \frac{M}{2\log_2(M)}\log(2),\quad
   \slopelog\BICM = \frac{2\log_2(M)}{M} \notag\;.
\end{align}
Noteworthy, the wideband slope is always positive.

In the case of natural labeling, due to the construction of the labeling, fixing the first bit position results in constrained constellations $\const_b^1$, which are equivalent to $D$-PPM for $b=0$ and to $D$-PPM with sign-inverted pulses for $b=1$ (cf., Table \ref{mapping:8biPPM}), yielding  $\mean{\const_b^1} = \pm2/M\cdot\ve{1}$.
For the other bit positions $\mu=2,\ldots,m$, the constrained constellations correspond to $(M/2 = D)$-biPPM (cf.,  Table \ref{mapping:8biPPM}).
After straight-forward calculations one obtains
\begin{align}
   \coneBICM = \frac{1}{D},\quad
   \ctwoBICM = &\frac{\log_2(2D)}{D} - \frac{1}{D}\left(1+\frac{1}{D}\right) \label{eq:coeffs_natbippm}\;.
\end{align}
In contrast to the (near-)Gray labeling, as $\ctwoBICM>0$ for $D\geq2$, the corresponding wideband slope is negative for all $M\geq4$ (cf.\ (\ref{eq:slopelog})).
However, in both cases the ratio $\EbNolim$ increases with increasing $M$.

\begin{table}[tb]
\caption{Binary labeling rules (near-Gray and natural) for $8$-biPPM.\vspace*{-4mm}}
\label{mapping:8biPPM}
\begin{center}
\begin{tabular}{l|rrrr|c|c}
\hline
   &\multicolumn{4}{c|}{signal element}&\multicolumn{2}{c}{binary label}\\
$i$&\multicolumn{4}{c|}{$\ve{x}$}      & near-Gray & natural  \\\hline
1  & [\hphantom{-}1 &0& 0& 0]         & 000       &000\\
2  & [\hphantom{-}0 &1& 0& 0]         & 001       &001\\
3  & [\hphantom{-}0 &0& 1& 0]         & 010       &010\\
4  & [\hphantom{-}0 &0& 0& 1]         & 011       &011\\
5  & [-1 &0& 0& 0]                    & 111       &100\\
6  & [\hphantom{-}0 &-1& 0& 0]        & 110       &101\\
7  & [\hphantom{-}0 &0 &-1& 0]        & 101       &110\\
8  & [\hphantom{-}0 &0 &0 &-1]        & 100       &111\\
\hline
\end{tabular}
\vspace*{-5mm}
\end{center}
\end{table}

\subsection{Noncoherent Detection}
\label{sec:noncoherent}
Due to the large signal bandwidth in IR-UWB systems, channel estimation required for coherent detection remains a challenging task.
Instead, noncoherent detection schemes are employed for IR-UWB, i.e., in particular, energy detection  in the case of $M$-PPM, and autocorrelation-based detection in the case of pulse-amplitude-modulated IR-UWB \cite{UWB:Witrisaletal:NoncoherentUWBSystems}.

Employing energy detection of $M$-PPM corresponds to computation of the decision metrics by elementwise squaring the components of the receive symbol $\ve{y}$, given in (\ref{eq:model}) \cite{book:Proakis:DigitalCommunications}.

In the case of one-dimensional signaling employing BPSK, noncoherent detection requires to differentially encode the information symbols in the phase transition of adjacent symbols.
Traditional symbolwise differential detection computes the decision metric for the $k$-th information symbol as
\begin{align}
   z_k = y_{k}\cdot y_{k-1} \label{eq:diffdetect}
\end{align}
where $y_k$ is the actual receive symbol and $y_{k-1}$ is the preceeding receive symbol, which serves as phase reference.
Eq.\,(\ref{eq:diffdetect}) can be seen as a (simplified) model for noncoherent detection of differential-transmitted reference IR-UWB based on a so-called autocorrelation receiver \cite{UWB:Win:IR}.

We restrict to a numerical evaluation of the CM and BICM capacity of the resulting nonlinear channel, but note that a detailed study of the CM capacity of energy detected orthogonal signals has been given in \cite{IT:Gursoy:EnergyEfficiencyofOrthogonalSignaling,IT:GursoyPoorVerdu:OnOffFSKforWidebandFading}, where it has been shown that $\EbNolim\CM\rightarrow\infty$ (and thus also $\EbNolim\BICM$) for all $M$.
A thorough (wideband) analysis of noncoherent detection of (bi-)orthogonal signals is beyond the scope of this paper.

\section{Numerical Results}
\label{sec:results}

\begin{figure}
\begin{center}
  \psfrag{xlabel}[ct]{\mbox{}\hspace*{45mm}$10\log_{10}(\Eb/\No)\,[\dB]$ $\rightarrow$}
  \psfrag{ylabel}[cb]{\mbox{}\hspace*{35mm}$C$ [bits per dimension] $\rightarrow$}
  \psfrag{M2}[rb]{$M=2$}
  \psfrag{M4}[rb]{$M=4$}
  \psfrag{M8}[rb]{$M=8$}
  \psfrag{M16}[rb]{$M=16$}
  \includegraphics[width= .9\columnwidth]{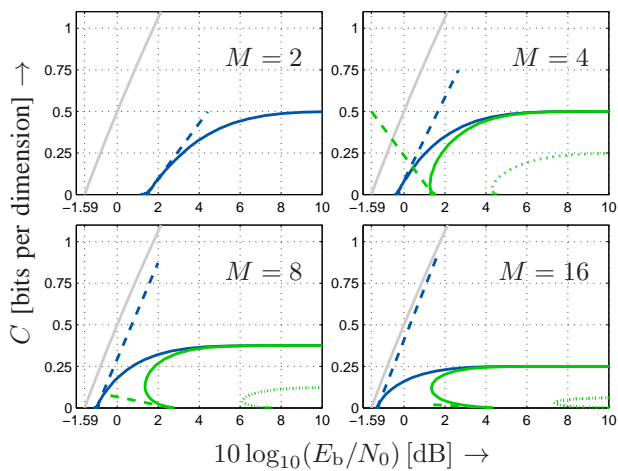}
   \caption{Capacity of coherent detection of $M$-PPM vs.\ $\Eb/\No\,[\dB]$. Blue: CM, green: BICM, dotted: bit level capacities, light-gray: Shannon capacity. $\EbNolim$ and wideband slope in bits/dim./$3\,\dB$ (dashed) indicated.}
   \label{fig:ppm}
\end{center}
\vspace*{-3mm}
\end{figure}

\begin{figure}
\begin{center}
  \psfrag{xlabel}[ct]{\mbox{}\hspace*{45mm}$10\log_{10}(\Eb/\No)\,[\dB]$ $\rightarrow$}
  \psfrag{ylabel}[cb]{\mbox{}\hspace*{35mm}$C$ [bits per dimension] $\rightarrow$}
  \psfrag{M2}[rb]{$M=2$}
  \psfrag{M4}[rb]{$M=4$}
  \psfrag{M8}[rb]{$M=8$}
  \psfrag{M16}[rb]{$M=16$}
  \includegraphics[width= .9\columnwidth]{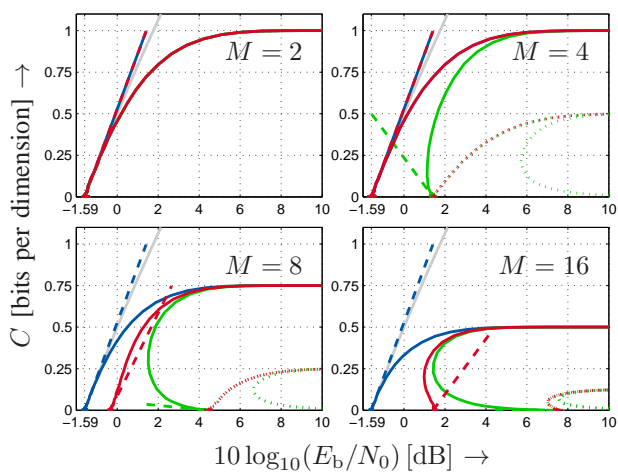}
   \caption{Capacity of coherent detection of $M$-biPPM vs.\ $\Eb/\No\,[\dB]$. Blue: CM, green: BICM with natural labeling, red: BICM with (near-)Gray labeling, dotted: bit level capacities, light-gray: Shannon capacity. $\EbNolim$ and wideband slope in bits/dim./$3\,\dB$ (dashed) indicated.}
   \label{fig:bippm}
\end{center}
\vspace*{-3mm}
\end{figure}

Fig.\,\ref{fig:ppm} and \ref{fig:bippm} depict the CM and BICM capacity (including the respective bit level capacities) in the case of coherent detection of $M$-PPM and $M$-biPPM, respectively, where $M=2,\,4,\,8,\,16$.
The capacities have been obtained from Monte-Carlo integration.
For comparison the Shannon capacity and the corresponding value $\EbNolim$ and the wideband slope (over a range of $3\,\dB$) are also shown and prove the results obtained in Sec.\,\ref{sec:BICMbiPPM}.
Especially for large $M$, a significant loss results from the restriction to BICM with parallel decoding in contrast to CM.
As expected, the bit level capacities of $M$-PPM are all equal.
In the case of $M$-biPPM with (near-)Gray labeling, all constrained constellations correspond to $M/2$-PPM.
Hence, also the bit level capacities are equal.
If natural labeling is applied, the bit level capacity of the bit specifying the pulse phase differs from the bit level capacities specifying the pulse position (whereas the remaining bit level capacities again correspond to PPM), which results in a significantly less BICM capacity.
Note, that in the case of BICM of $8$- and $16$-biPPM and (near-)Gray labeling, the depicted wideband slopes point out the limitations of a first order analysis.

\begin{figure}
\begin{center}
  \psfrag{xlabel}[ct]{\mbox{}\hspace*{45mm}$10\log_{10}(\Eb/\No)\,[\dB]$ $\rightarrow$}
  \psfrag{ylabel}[cb]{\mbox{}\hspace*{35mm}$C$ [bits per dimension] $\rightarrow$}
  \psfrag{M2}[rb]{$M=2$}
  \psfrag{M4}[rb]{$M=4$}
  \psfrag{M8}[rb]{$M=8$}
  \psfrag{M16}[rb]{$M=16$}
  \includegraphics[width= .9\columnwidth]{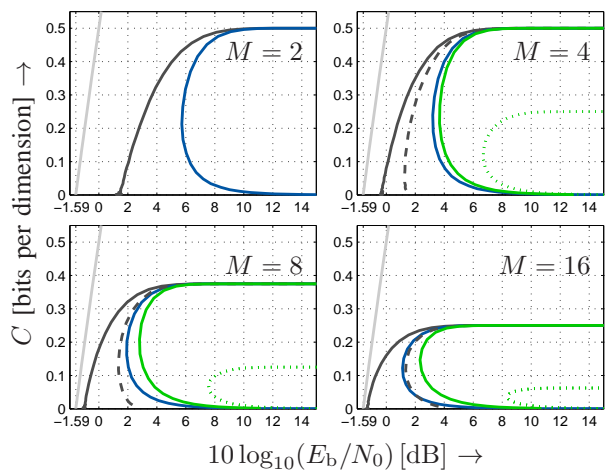}
   \caption{Capacity of energy detection of $M$-PPM vs.\ $\Eb/\No\,[\dB]$. Blue: CM, green: BICM, dotted: bit level capacities, dark-gray: CM and BICM capacity of coherent detection, light-gray: Shannon capacity.}
   \label{fig:ppm_ed}
\end{center}
\vspace*{-3mm}
\end{figure}

\begin{figure}
\begin{center}
  \psfrag{xlabel}[ct]{\mbox{}\hspace*{0mm}$10\log_{10}(\Eb/\No)\,[\dB]$ $\rightarrow$}
  \psfrag{ylabel}[cb]{\mbox{}\hspace*{0mm}$C$ [bits per dimension] $\rightarrow$}
  \psfrag{M2}[rb]{$M=2$}
  \psfrag{M4}[rb]{$M=4$}
  \psfrag{M8}[rb]{$M=8$}
  \psfrag{M16}[rb]{$M=16$}
  \includegraphics[width= .6\columnwidth]{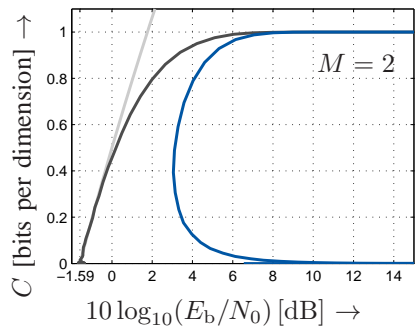}
   \caption{Capacity of differential detection of BPSK vs.\ $\Eb/\No\,[\dB]$. Blue: CM and BICM, dark-gray: CM capacity of coherent detection, light-gray: Shannon capacity.}
   \label{fig:bpsk_acr}
\end{center}
\vspace*{-6mm}
\end{figure}

Fig.\,\ref{fig:ppm_ed} and \ref{fig:bpsk_acr} depict the CM and BICM capacity of energy detection of $M$-PPM, and differential detection of BPSK, respectively.
It can be observed that the loss induced by using BICM instead of CM is less compared to the coherent case.
However, as known from \cite{IT:Gursoy:EnergyEfficiencyofOrthogonalSignaling}, in both cases the ratio $\EbNolim$ tends to infinity at a very small and negative wideband slope for both, PPM and BPSK.
Hence, for the noncoherent schemes, but also for most coherent BICM schemes, 
the minimum ratio of $\Eb/\No$ is obtained at nonzero capacity.
An analytical expression for this point remains an open issue.

\section{Conclusions and Design Rules}
\label{sec:conclusions}
%

Motivated by IR-UWB as a possible application, we have analyzed the BICM capacity using (bi-)orthogonal signals and put particular focus on the wideband regime.
We have specified a binary labeling rule for biPPM, which performs significantly better compared to the natural labeling rule.
The gained insights emphasize that in the design of IR-UWB systems, the rates at the operating point have to be carefully selected.


The presented analysis is based on optimum channel coding and assesses ultimate capacity limits.
The shortcomings of applicable channel coding and a more detailed system model of IR-UWB have to be taken into account in future works assessing realistic IR-UWB system design.

%
%
%
%



\clearpage

\appendix[Derivation of Coefficients $\cone$ and $\ctwo$]
In this appendix, a detailed derivation of the coefficients $\cone$ and $\ctwo$ is given, which are 
required for the wideband analysis 
 of BICM and CM using $M$-PPM and $M$-biPPM (cf.\ (\ref{eq:ebnolim}), (\ref{eq:slopelog})).
For BICM using $M$-biPPM, (near-)Gray and natural labeling are distinguished.
In the case of CM, this requires to calculate the diagonal elements of $\cov{\const}$ and $\cov{\const}^2$ (cf.\ (\ref{eq:c2_cm})).
In the case of BICM, the covariance matrices of the constrained constellations are also required (cf.\ (\ref{eq:c1_bicm}) and (\ref{eq:c2_bicm})).
Using these coefficients, the ratio $\EbNolim$ and the wideband slope $\slopelog$ can directly be computed for the respective signaling schemes.
The resulting values are summarized in Table\,\ref{table:values}.

\subsection{$M$-PPM CM}
From (\ref{eq:mean}), we have $\mean{\const} = 1/M\cdot \ve{1}$ and 
\begin{align*}
   \|\mean{\const}\|^2 &= 1/M\\
    \mean{\const} \trans \mean{\const} &= \frac{1}{M^2} \ve{1}\trans\ve{1}\\
   \E\{\ve{X}\trans\ve{X}\} &= \ve{I}/M
\end{align*}
Thus, using (\ref{eq:c1_cm}) and (\ref{eq:c2_cm}) and (\ref{eq:ebnolim}), (\ref{eq:slopelog}), we have
\begin{align*}
\coneCM &=  1 - 1/M\\
\ctwoCM &= -1/M\cdot (1 - 1/M)\\
\EbNolim\CM &= \frac{M}{M-1} \log(2)\\
\slopelog\CM &= -\frac{(1 - 1/M)^2}{1/M\cdot (1 - 1/M) \cdot D} = \frac{M-1}{M}
\end{align*}

\subsection{$M$-PPM BICM}
Since all constrained constellation correspond to $(M/2)$-PPM, we have for all $\mu=1,\ldots,m$, $b\in\{0,1\}$:
\begin{align*}
\trace(\cov{\const_b^{\mu}}) &= 1 - 2/M\\
\trace(\cov{\const_b^{\mu}}^2) &= 2/M\cdot (1 - 2/M)
\end{align*}
Using these results, from (\ref{eq:c1_bicm}) and (\ref{eq:c2_bicm}) the coefficients for BICM are given as
\begin{align*}
\coneBICM &= \log_2(M) \left((1 - 1/M) - (1 - 2/M)\right)\\
      &= \log_2(M) / M\\
\ctwoBICM &= \log_2(M) \left( -1/M\cdot (1 - 1/M) + 2/M\cdot (1 - 2/M)  \right)\\
      &= \log_2(M) / M \left(1 - 3/M\right)
\end{align*}
which results in
\begin{align*}
   \EbNolim\BICM &= \frac{M}{\log_2(M)} \log(2)\\
   \slopelog\BICM &= -\frac{\log_2(M)}{M(M-3)}
\end{align*}
Consequently, the wideband slope is negative for all $M\geq4$.

\begin{table}[tb]
\caption{Ratio $\EbNolim$ and wideband slope for $M$-(bi)PPM in the case of CM and BICM}
\label{table:values}
\begin{center}
\begin{tabular}{llc|c|c|c|c}
&&$M$&\multicolumn{2}{c|}{$\EbNolim\,[\dB]$}&\multicolumn{2}{c}{$\slopelog\,\frac{\mathrm{bit/dim.}}{3\,\dB}$}\\\hline
\multirow{10}{*}{\rotatebox{90}{PPM}}&
\multirow{5}{*}{\rotatebox{90}{CM}}
    &$2$&\multicolumn{2}{c|}{\hphantom{-}1.42}&\multicolumn{2}{c}{0.5}\\
    &&$4$&\multicolumn{2}{c|}{-0.34}&\multicolumn{2}{c}{0.75}\\
    &&$8$&\multicolumn{2}{c|}{-1.01}&\multicolumn{2}{c}{0.875}\\
    &&$16$&\multicolumn{2}{c|}{-1.31}&\multicolumn{2}{c}{0.9375}\\
    &&$32$&\multicolumn{2}{c|}{-1.45}&\multicolumn{2}{c}{0.9688}\\
\cline{2-7}&
\multirow{5}{*}{\rotatebox{90}{BICM}}
  &$2$&\multicolumn{2}{c|}{\hphantom{-}1.42}&\multicolumn{2}{c}{1}\\
  &&$4$&\multicolumn{2}{c|}{\hphantom{-}1.42}&\multicolumn{2}{c}{-0.5}\\
  &&$8$&\multicolumn{2}{c|}{\hphantom{-}2.67}&\multicolumn{2}{c}{-0.075}\\
  &&$16$&\multicolumn{2}{c|}{\hphantom{-}4.43}&\multicolumn{2}{c}{-0.0192}\\
  &&$32$&\multicolumn{2}{c|}{\hphantom{-}6.47}&\multicolumn{2}{c}{-0.0054}\\
\hline\hline
\multirow{7}{*}{\rotatebox{90}{biPPM}}&
\multirow{2}{*}{\rotatebox{90}{CM}}
  &\raisebox{-2mm}{$\forall M$}&\multicolumn{2}{c|}{\raisebox{-2mm}{-1.59}}&\multicolumn{2}{c}{\raisebox{-2mm}{1}}\\[3mm]
\cline{2-7}&
&&natural&(near-)Gray&natural&(near-)Gray\\
\cline{4-7}&\multirow{4}{*}{\rotatebox{90}{BICM}}
    &$2$&-1.59&-1.59&1.0&1.0\\
    &&$4$&\hphantom{-}1.42&-1.59&-0.5&1.0\\
    &&$8$&\hphantom{-}4.42&-0.34&-0.04&0.75\\
    &&$16$&\hphantom{-}7.43&\hphantom{-}1.42&-0.005&0.5\\
    &&$32$&\hphantom{-}10.45&\hphantom{-}3.46&-0.001&0.3125\\

\end{tabular}
\end{center}
\end{table}

\subsection{$M$-biPPM CM}
For biorthogonal constellations $\mean{\const} = \ve{0}$, thus, we have
\begin{align*}
\E\{\ve{X}\trans\ve{X}\} &= \ve{I}/D
\end{align*}
Consequently,
\begin{align*}
   \coneCM &= 1\\
   \ctwoCM &= -\trace((\ve{I}/D)^2)=-D/D^2 = -1/D\\
\EbNolim\CM &=  \log(2)\\
\slopelog\CM &= -\frac{1}{D\cdot(-1/D)} = 1
\end{align*}

\subsection{$M$-biPPM BICM}
The BICM capacity of $M$-biPPM depends on the applied binary labeling.

\subsubsection{(Near-)Gray Labeling}
In the case of (near-)Gray labeling, all constrained constellations correspond to $(D=M/2)$-PPM (cf.\ Table\,\ref{mapping:8biPPM}), thus, we have  for $\mu=1,\ldots,m$, $b\in\{0,1\}$:
\begin{align*}
\trace(\cov{\const_b^{\mu}}) &= 1-1/D\\
\trace(\cov{\const_b^{\mu}}^2)&=1/D(1-1/D)
\end{align*}
The coefficients compute to
\begin{align*}
\coneBICM &= \log_2(2D)/D\\
\ctwoBICM &= \log_2(2D)/D (-1/D)
\end{align*}
Hence,
\begin{align*}
   \EbNolim\BICM &= \frac{M}{2\log_2(M)}\log(2)\\
   \slopelog\BICM &= \frac{2\log_2(M)}{M}
\end{align*}

\subsubsection{Natural Labeling}
In the case of natural labeling, the constrained constellation $\const_b^1$, with fixed first bit position corresponds to $(D=M/2)$-PPM.
Fixing the remaining bit positions, the constrained constellations correspond to $(D=M/2)$-biPPM. Consequently,  $i,k=1,\ldots,M$, $i\neq k$, $\mu=2,\ldots,m$, $b\in\{0,1\}$,
\begin{align*}
\trace(\cov{\const_b^1}) &= 1-1/D\\
\trace(\cov{\const_b^1}^2)&=1/D(1-1/D) \\
\trace(\cov{\const_b^{\mu}}) &= 1\\
\trace(\cov{\const_b^{\mu}}^2)&=1/(D/2)
\end{align*}
which gives
\begin{align*}
\coneBICM &= m - (1-1/D) - (m-1)\cdot 1 = 1/D\\
\ctwoBICM &= m\cdot(-1/D) + 1/D(1-1/D) + (m-1) \cdot2/D\\
      &= m/D - 1/D (1+1/D)
\end{align*}
Finally,
\begin{align*}
   \EbNolim\BICM &= D\log(2) = \frac{M}{2}\log(2)\\
   \slopelog\BICM &= -\frac{1}{D\left(D\log_2(2D) - D +1\right)}\\
             &= -\frac{4}{M\left(M\log_2(M) - M +2\right)}
\end{align*}

\appendix[The Wideband Slope]

Using (15), (35), and (142) of \cite{IT:Verdu:SpectralEfficiency}, and the definitions
\begin{align*}
   \dot{C} &\deq \left.\frac{\dx C(\Es/\No)}{\dx\Es/\No}\right|_{\Es/\No=0}\\
\ddot{C} &\deq \left.\frac{\dx^2 C(\Es/\No)}{(\dx\Es/\No)^2}\right|_{\Es/\No=0}
\end{align*}
computed in nats, the wideband slope in linear scale reads (neglecting second- and higher-order terms)

\begin{align*}\allowdisplaybreaks
   \frac{C(\Eb/\No)}{\Eb/\No - \EbNolim } =\hspace*{-2cm}&\\
&\quad\frac{\dot{C}\log_2(\e)\cdot\Es/\No}{\left(\dot{C}\log_2(\e) + \frac{\ddot{C}}{2}\Es/\No \log_2(\e)\right)^{-1} - \frac{\log(2)}{\dot{C}}}\\
  =& \frac{\dot{C}^2\Es/\No\log_2^2(\e)}{\left(1+\frac{\ddot{C}}{2\dot{C}}\Es/\No\right)^{-1} -1}\\
  =& \frac{\dot{C}^2\Es/\No\log_2^2(\e) \left(1+\frac{\ddot{C}}{2\dot{C}}\Es/\No\right) }{1 -1 - \frac{\ddot{C}}{2\dot{C}}\Es/\No}\\
  =& -2\frac{\dot{C}^3}{\ddot{C}}\log_2^2(\e) - \dot{C}^2\Es/\No\log_2^2(\e)
\intertext{and for $\Eb/\No \rightarrow \EbNolim$, i.e., $\Es/\No=0$,  we have}
  =& -2\frac{\dot{C}^3}{\ddot{C}}\log_2^2(\e) \deq \slopelin
\end{align*}
Applying the definitions $\cone \deq \dot{C}$ and $\ctwo \deq \ddot{C}/2$, thus using the notation of \cite{BICM:MartinezFabregasCaireWillems:BICMWideband}, yields
\begin{align*}
   \slopelin &=-\frac{\cone^3}{\ctwo}\log_2^2(\e) = -\frac{\cone^3}{\ctwo\log^2(2)}
\end{align*}

The conversion from linear scale to semilogarithmic scale is obtained by using the following relation.
The slope $\slopelin = g'(x_0)$ of the tangent at the point $x_0$ of a function $g(x)$ converts to a slope $S_0$ of the tangent at the point $10\log_{10}(x_0)$ of the function $g(10\log_{10}(x))$ via
\begin{align*}
   S_0 = \frac{\log(10)}{10} x_0 \slopelin
\end{align*}
Here, we have $x_0 = \EbNolim = \log(2) / \cone$, thus
\begin{align*}
   S_0 &= -\frac{\log(10)}{10} \frac{\log(2)}{\cone} \frac{\cone^3}{\ctwo\log^2(2)}\\
             &= -\frac{\cone^2}{\ctwo} \cdot\frac{1}{10\log_{10}(2)}
\end{align*}
which gives $\slopelog = -\frac{\cone^2}{\ctwo}$ as the wideband slope in bits per $3\,\dB$.

\end{document}